\documentclass{article}

\usepackage{arxiv}
\usepackage{cite}
\usepackage[utf8]{inputenc} 
\usepackage[T1]{fontenc}    
\usepackage{hyperref}       
\usepackage{url}            
\usepackage{booktabs}       
\usepackage{amsfonts}       
\usepackage{nicefrac}       
\usepackage{microtype}      
\usepackage{graphicx}
\usepackage[sort,compress,square,numbers]{natbib}
\usepackage{doi}
\usepackage[nottoc]{tocbibind}
\usepackage{float}
\usepackage{siunitx}
\usepackage{soul}
\usepackage{xcolor}
\usepackage[normalem]{ulem}

\title{Increasing the Field-of-View Radiation Efficiency of \\Optical Phased Antenna Arrays}

\author{ \href{https://orcid.org/0000-0001-7730-3489}{\includegraphics[scale=0.06]{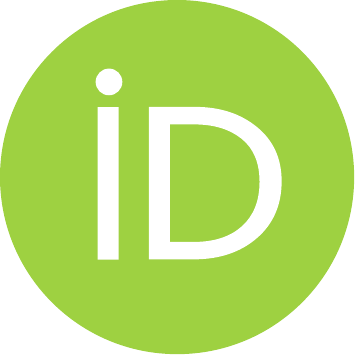}\hspace{1mm}Henna Farheen}\\
	Paderborn University\\
	Theoretical Electrical Engineering\\
	Warburger Str. 100, 33098 Paderborn, Germany \\
	\texttt{henna18@campus.uni-paderborn.de} \\
	\And
    \hspace{1mm}Andreas Strauch \\
	Paderborn University\\
	Theoretical Electrical Engineering\\
	Warburger Str. 100, 33098 Paderborn, Germany \\
	\texttt{astrauch@campus.uni-paderborn.de} \\
	\And
	\href{https://orcid.org/0000-0002-5950-6618}{\includegraphics[scale=0.06]{orcid.pdf}\hspace{1mm}J. Christoph Scheytt}\\
	Paderborn University\\
	System and Circuit Technology\\
	F\"{u}rstenallee 11, 33102 Paderborn, Germany\\
	\texttt{christoph.scheytt@hni.uni-paderborn.de} \\
	\And
	\href{https://orcid.org/0000-0001-6431-746X}{\includegraphics[scale=0.06]{orcid.pdf}\hspace{1mm}Viktor Myroshnychenko} \\
    Paderborn University\\
	Theoretical Electrical Engineering\\
	Warburger Str. 100, 33098 Paderborn, Germany \\
	\texttt{viktor.myroshnychenko@uni-paderborn.de} \\
	\And
	\href{https://orcid.org/0000-0001-7059-9862}{\includegraphics[scale=0.06]{orcid.pdf}\hspace{1mm}Jens F\"orstner} \\
    Paderborn University\\
	Theoretical Electrical Engineering\\
	Warburger Str. 100, 33098 Paderborn, Germany \\
	\texttt{jens.foerstner@uni-paderborn.de} \\
}

\date{}

\renewcommand{\headeright}{}
\renewcommand{\undertitle}{}
\renewcommand{\shorttitle}{Increasing the Field-of-View Radiation Efficiency of Optical Phased Antenna Arrays}

\hypersetup{
pdftitle={Increasing the Field-of-View Radiation Efficiency of Optical Phased Antenna Arrays},
pdfkeywords={Dielectric antenna, phased arrays, field of view, directivity},
}

\begin{document}
\maketitle

	\begin{abstract}
		Silicon photonics in conjunction with complementary metal-oxide-semiconductor (CMOS) fabrication has greatly enhanced the development of integrated optical phased arrays. This facilitates a dynamic control of light in a compact form factor that enables the synthesis of arbitrary complex wavefronts in the infrared spectrum. We numerically demonstrate a large-scale two dimensional silicon-based optical phased array (OPA) composed of nanoantennas with circular gratings that are balanced in power and aligned in phase, required for producing elegant radiation patterns in the far-field. For a wavelength of \SI[mode=text]{1.55}{\micro\metre}, we optmize two antennas for the OPA exhibting an upward radiation efficiency as high as $90\%$, with almost 6.8\% of optical power concentrated in the field of view. Additionally, we believe that the proposed OPAs can be easily fabricated and would have the ability of generating complex holographic images, rendering them an attractive candidate for a wide range of applications like LiDAR sensors, optical trapping, optogenetic stimulation and augmented-reality displays.
	\end{abstract}
\section{Introduction}
	Phased arrays are devices which consist of multiple receiving or transmitting antennas \cite{balanis2015antenna}, where manipulation of amplitude and phase of each antenna element can be used to control the shape and direction of radiation field from the array, utilizing constructive and destructive interference \cite{ashtiani2019n}. For decades, the radio-frequency phased arrays were rigorously investigated and extensively employed in a variety of applications going from the radar to broadcasting applications \cite{fenn2000development}. However, their large-scale deployment was difficult and came with large manufacturing costs, making their optical counterparts more attractive which could offer on-chip integration of thousands of elements for short optical wavelengths of interest at a much lower cost and form factor\cite{sun2013large,benedikovivc2022circular}. These optical analogues are versatile devices which could be utilized for several applications like free-space communication \cite{rabinovich2015free,neubert1994experimental}, optical switches \cite{blanche2017diffraction}, holographic displays \cite{zhou2015design}, and light detection and ranging (LiDAR) \cite{poulton2019long,levinson2011towards,poulton2017coherent,bhargava2019fully}.
	
	In this regard, a promising platform to furnish high-yield, cost efficient commercial systems is silicon photonics, which is highly compatible with the standard CMOS technology \cite{chung2017monolithically}. This is commonly accomplished using the silicon-on-insulator (SOI) process, where the top silicon layer realizes all the active and passive components of the array like the directional couplers, phase shifters, waveguides and optical antennas \cite{abediasl2015monolithic}. A plethora of optical phased arrays (OPAs) have already been developed using this combination \cite{van2010two,van2009off,doylend2011two,doylend2012hybrid}. Silicon photonics makes it possible to reliably integrate large number of densely packed microelectronics. The number of these elements and the spacing between them can be used to tailor the angular resolution, beamwidth, grating lobe spacing and power of the emission pattern \cite{fatemi2019nonuniform}. In particular, a high resolution can be obtained by increasing the number of optical antennas, but on the other hand, their large size results in an inter-element spacing that is larger than the optical wavelength $\lambda$ \cite{sun2013large}. As a consequence, an increased number of grating lobes appear undesirably which limits the field of view (FOV) of the OPA and in turn its steering range \cite{fatemi2020breaking}. Furthermore, demonstrations have also been made to realize OPAs in other integrated photonic platforms like indium phosphide \cite{guo2013two}, silicon nitride \cite{sun2021parallel}, III/V hybrid platforms \cite{hulme2015fully,doylend2012hybrid}, liquid crystals \cite{resler1996high}, etc. \cite{mcmanamon1996optical}.  
	\begin{figure}[b!]
		\centering
		\includegraphics[width=\linewidth]{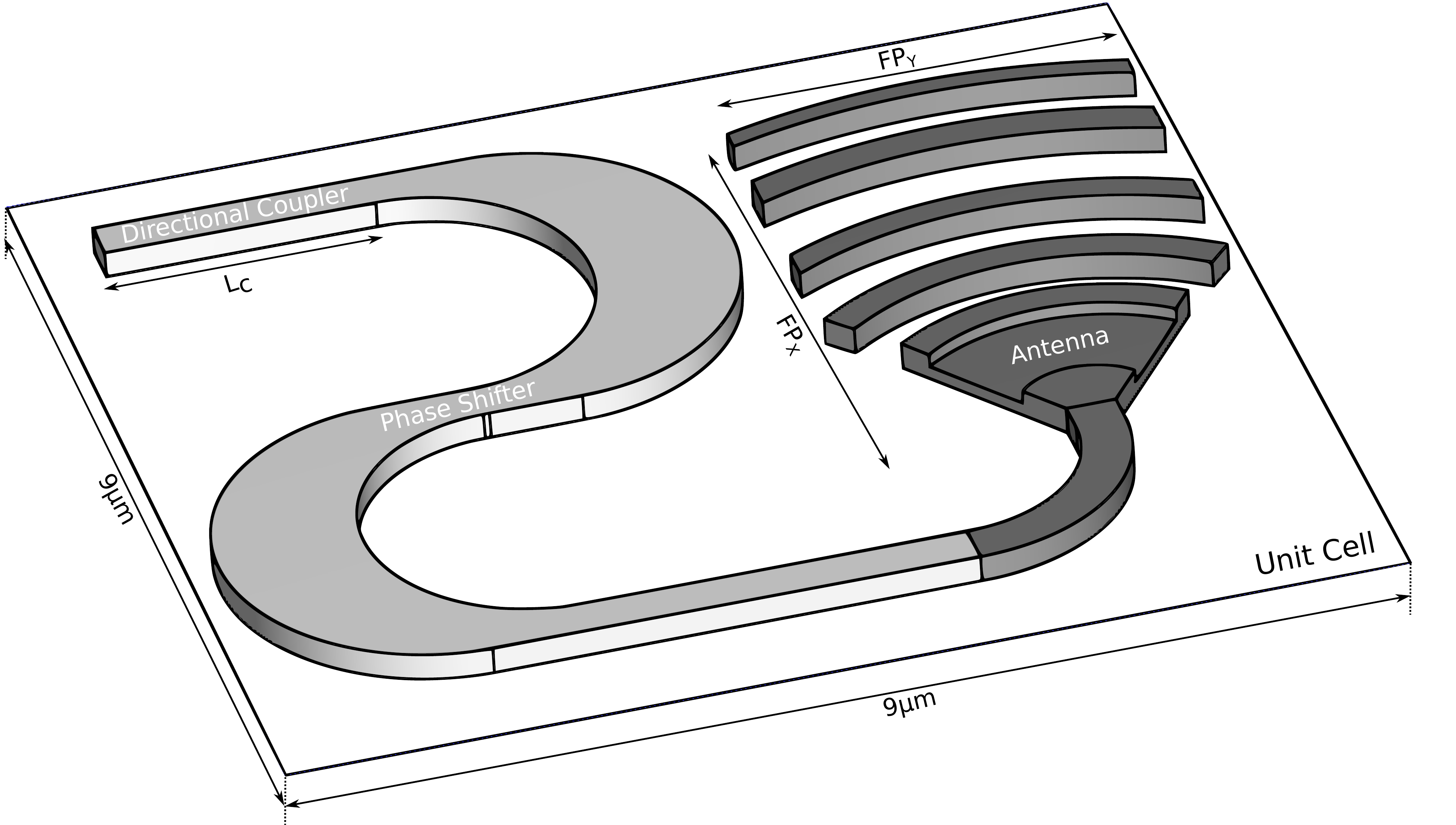}
		\caption{Schematic representation of the \SI[mode=text]{9}{\micro\metre}$\times$\SI[mode=text]{9}{\micro\metre} unit cell. The three main components include the direction coupler, phase shifter and the dielectric antenna. The parameters FP$_\mathrm{x}$ and FP$_\mathrm{y}$ represent the footprint of the antenna along the $x$- and $y$- coordinates, respectively. The dark grey region shows the part of the structure optimized in this work.}
		\label{fig:figure1}
	\end{figure}
	In this work, we aim at improving the efficiency in the FOV for the OPA presented in Ref. \cite{sun2013large1}. In particular, we use full-wave numerical simulations in conjunction with particle swarm optimization (PSO) to optimize the amount of power directed by the dielectric nanoantenna from a unit cell into the FOV \cite{farheen2022broadband,farheen2022optimization,leuteritz2021dielectric}. We use the technical specifications of each component in the unit cell to design an optimized antenna which fits exactly the same compact footprint of \SI[mode=text]{9}{\micro\metre}$\times$\SI[mode=text]{9}{\micro\metre} in Ref. \cite{sun2013large1}. Our optimized antenna directs $4.5$ times more power into the FOV than the reference antenna. Furthermore, we propose a modified structure that can further enhance this power to $10$ folds by using a simple, yet efficient Bragg reflector made of silicon. We anticipate that our optimized antennas can be easily fabricated and utilized for enhancing the performance of LiDARs and OPAs used in different applications.    
	
	\section{Numerical Setup}\label{sec:NS}
	
	\begin{figure*}[t!]
		\centering
		\includegraphics[width=\textwidth]{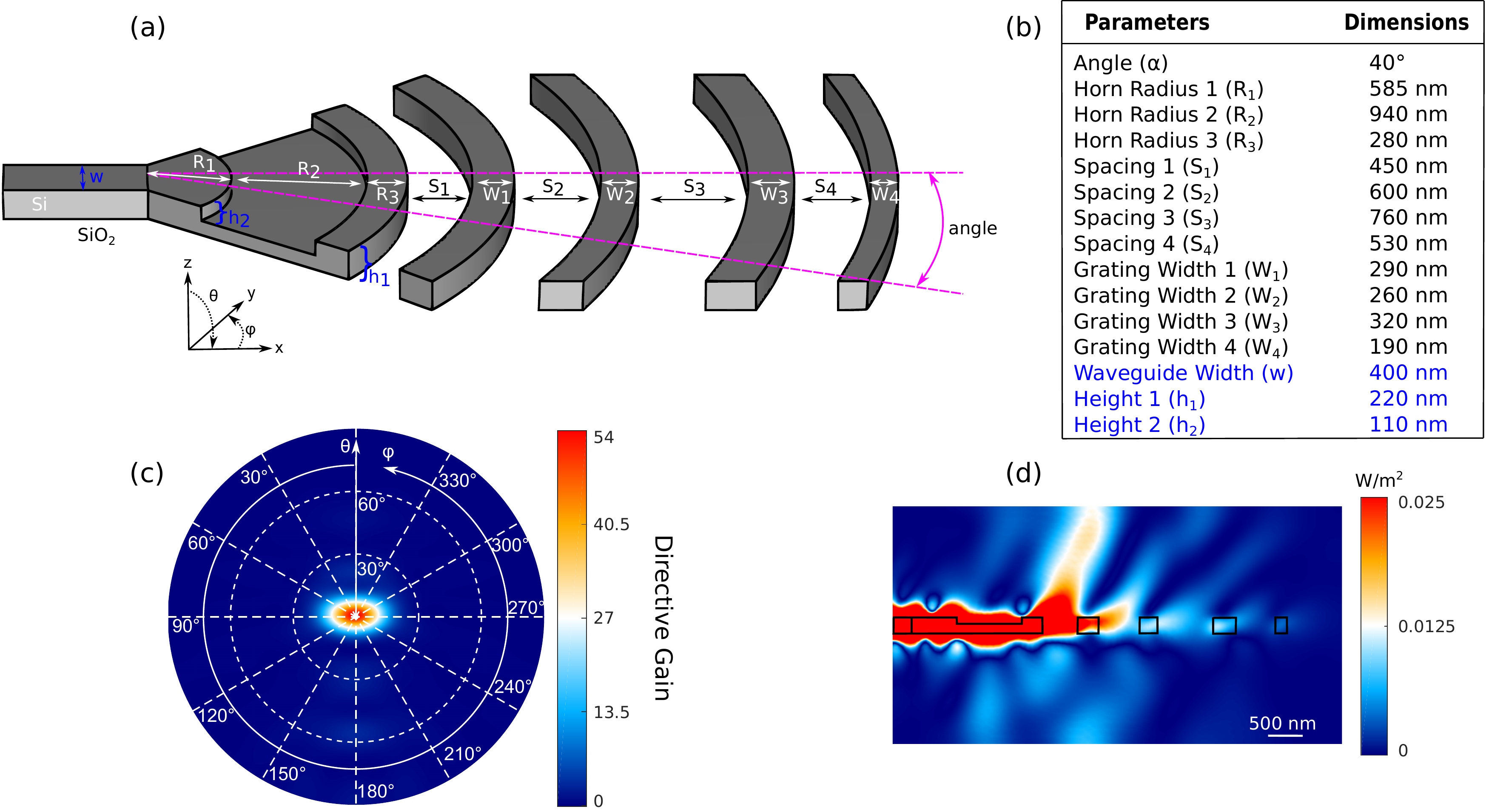}
		\caption{(a) Schematic representation of the optimized dielectric antenna, highlighting the parameters used in the optimization. (b) Design parameters obtained for the optimized antenna. The parameters depicted in blue were fixed as constant values. (c) Calculated angular linear directive gain distribution of the optimized antenna exhibiting a directivity of $D=54$ at $\theta=0^{\circ}$. (d) Calculated near-field distribution of the power flow for the optimized antenna in the $xz$-plane at $y=0$.}
		\label{fig:figure2}
	\end{figure*}
	
	In this work, we optimize the emission characteristics of two different antennas that can be used in a two-dimensional OPA. Each unit cell of the OPA comprises of three main components, namely, directional coupler (DC), phase shifter (PS) and a low-loss dielectric antenna as shown in Fig.~\ref{fig:figure1}. The technical specifications of the DC and PS were taken into consideration for defining the footprint constraints in the optimization of the antenna component, in order to fit it exactly in the same compact footprint of \SI[mode=text]{9}{\micro\metre}$\times$\SI[mode=text]{9}{\micro\metre} as the original design in \cite{sun2013large1}. The antennas are made of silicon with a refractive index of $\mathrm{n_{Si}=3.48}$ and are simulated in a silicon-dioxide cladding having a refractive index of $\mathrm{n_{SiO_2}=1.45}$. The large refractive index contrast between the two materials facilitates a stronger light-matter interaction, which in turn helps in obtaining higher radiation efficiencies from the antennas. Fig.~\ref{fig:figure2}a and Fig.~\ref{fig:figure3}a show the schematic representation of the antennas, which are oriented along the $xy$-plane with a direction of propagation along the $x$-axis. These antennas are excited with the fundamental transverse electric (TE) mode at an operational wavelength of \SI[mode=text]{1.55}{\micro\metre}. Full-wave numerical simulations were performed using CST Microwave Studio in frequency domain utilizing the finite element method \cite{CST}. The computational domain of the simulations is surrounded with an open boundary condition in which the far-field for the antenna is obtained using near-to-far-field transformation utilizing the fields at the bounding box of the simulation domain \cite{taflove2005computational}. The optimization utilized the particle swarm algorithm, a heuristic global optimizer in conjunction with the trust-region method, which is a local optimization technique. We optimize the power efficiency of the antenna in the FOV, i.e. the grating lobes free region that along $\theta$ and $\varphi$ can be defined as \cite{palmer2020diffraction,chung2017monolithically}
	\begin{equation}
		- \mathrm{\sin^{-1}\left(\frac{\lambda_n}{2d}\right)}<\Delta \text{FOV} < \mathrm{\sin^{-1}\left(\frac{\lambda_n}{2d}\right)},
	\end{equation}
	where $\mathrm{d}$ is the size of the unit cell, which is \SI[mode=text]{9}{\micro\metre} in this case and $\mathrm{\lambda_n}$ is the wavelength in the medium of propagation. Therefore, for such an array configuration the FOV can be approximated to $6.8^\circ \times 6.8^\circ$ along $\theta$ and $\varphi$, respectively with $m$ propagating interference orders, where $m$ is the largest value that satisfies $\mid m\lambda_n/d \mid <2$. For our unit cell, we have 16 interference orders in each direction.
	
	\section{Results and Discussion}  
	
	\begin{figure*}[t!]
		\centering
		\includegraphics[width=\textwidth]{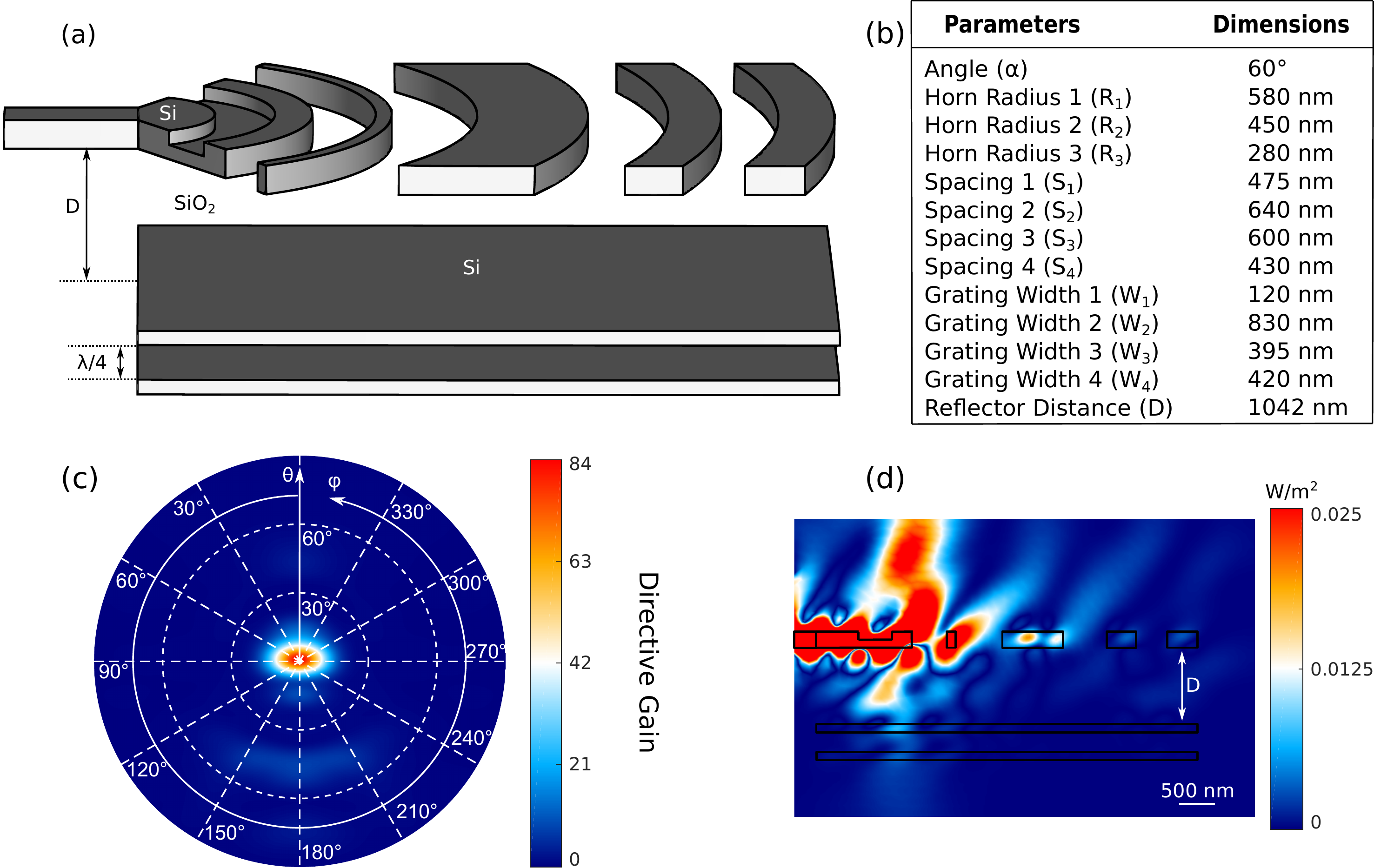}
		\caption{(a) Schematic representation of the optimized dielectric antenna with the Bragg reflector, highlighting the parameters used in the optimization. (b) Design parameters obtained from the optimization process. (c) Calculated angular linear directive gain distribution of the antenna exhibiting a directivity of $D=84$ at $\theta=1^{\circ}$. (d) Calculated near-field distribution of the power flow for the optimized antenna in the $xz$-plane at $y=0$.}
		\label{fig:figure3}
	\end{figure*}
	
\begin{figure*}[t!]
		\centering
		\includegraphics[width=\textwidth]{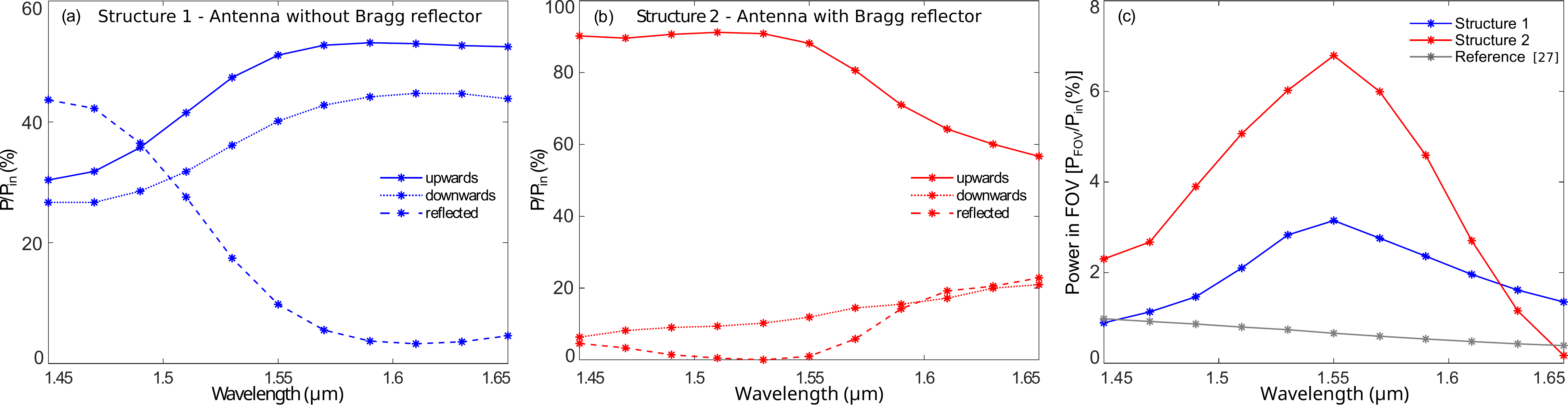}
		\caption{Calculated optical radiation efficiencies of optimized antenna (a) without and (b) with the Bragg reflector. (c) Calculated power in the FOV for both optimized antennas in comparison to the antenna presented in Ref.\cite{sun2013large1}.}
		\label{fig:figure4}
	\end{figure*}

	We begin with the optimization of the first structure shown in Fig.~\ref{fig:figure2}a. It comprises of an initial waveguide which guides light into the horn-section with a partial etch followed by five circular gratings with a variable pitch. This gives twelve optimization parameters as highlighted in the figure. The parameters include the angle ($\alpha$), horn radius 1 (R$_1$), horn radius 2 (R$_2$), horn radius 3 (R$_3$), spacing 1 (S$_1$), spacing 2 (S$_2$), spacing 3 (S$_3$), spacing 4 (S$_4$), grating width 1 (W$_1$), grating width 2 (W$_2$), grating width 3 (W$_3$) and grating width 4 (W$_4$). The table in Fig.~\ref{fig:figure2}b highlights the values of these parameters for the optimized antenna and, additionally, the values in blue specify the fixed parameters for the excitation waveguide width (w), the full height (h$_1$), and partial etch height (h$_2$) of the antenna. The partial etch helps in breaking the up-down symmetry of the antenna in order to have a higher upward radiation efficiency in comparison to the downward radiation efficiency, utilizing the constructive-destructive interference \cite{roelkens2006high,fan2007high}. The optimized antenna exhibits a directivity of 54 directed at an angle of $\theta=0^\circ$ and $\varphi=0^\circ$, as shown in Fig.~\ref{fig:figure2}c, representing the calculated linear directive gain distribution of the antenna. Fig.~\ref{fig:figure2}d shows the near-field power distribution of the structure which exhibits upwards propagation of the power. Taking a look at the broadband efficiencies of this structure, at a wavelength of $\mathrm{\lambda}$\SI[mode=text]{=1.55}{\micro\metre}, this structure has an upward efficiency of 51\%, downward efficiency of 39\%, and reflection efficiency of 10\% back to the waveguide (see Fig.~\ref{fig:figure4}a). Interestingly, Fig.~\ref{fig:figure4}c reveals that approximately 3.2\% of the optical power is radiated into the FOV. Overall, the antenna structure has a compact footprint of approximately \SI[mode=text]{3.27}{\micro\metre}$\times$\SI[mode=text]{5.2}{\micro\metre}.

	\begin{figure*}[b!]
		\centering
		\includegraphics[width=\textwidth]{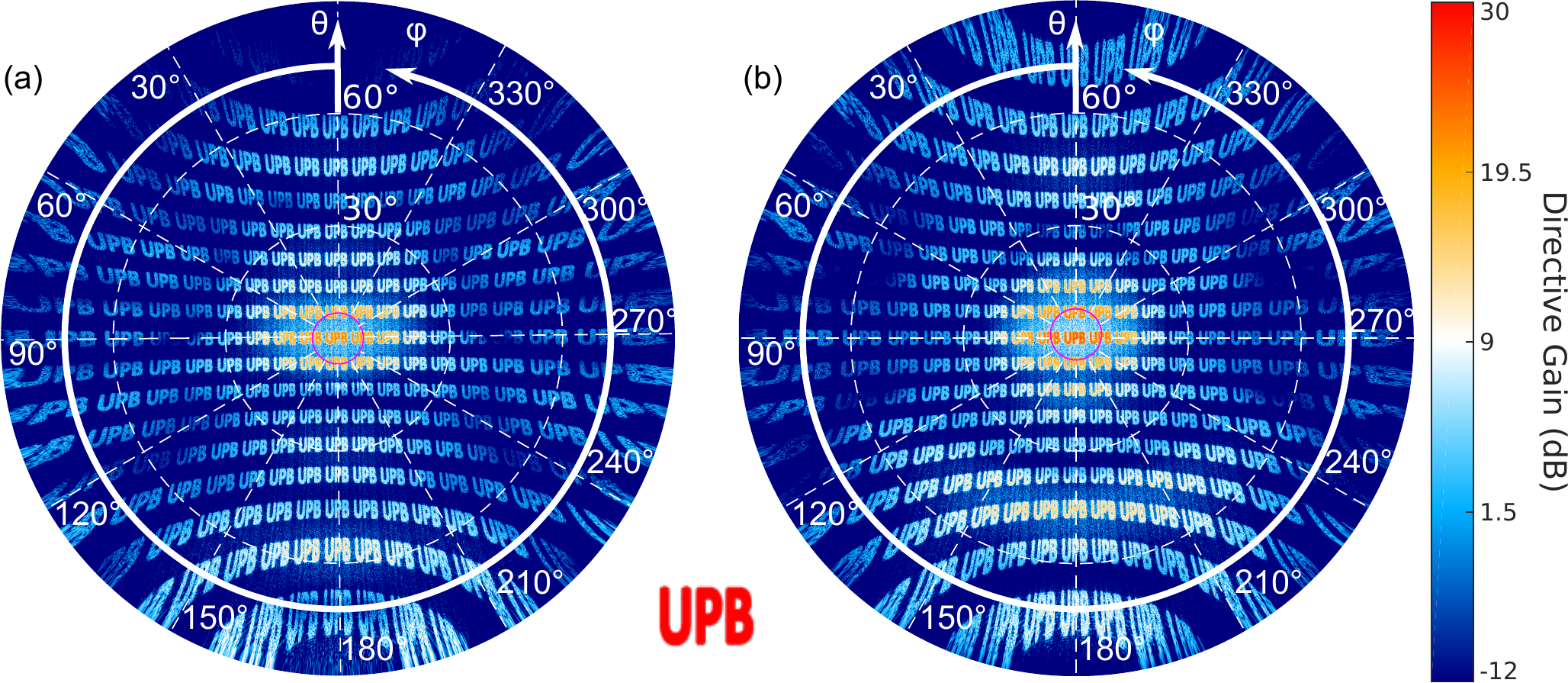}
		\caption{Calculated far-field radiation patterns for a $64 \times 64$ phased array configuration with respect to the optical antenna (a) without and (b)with the Bragg reflector to generate the Paderborn University logo (UPB) . The original logo used in the process is shown in the middle. The pink circle highlights the FOV region.}
		\label{fig:figure5}
	\end{figure*}
	
	In order to further improve the upward efficiency, we introduce a silicon-based Bragg reflector in our structure. The reflector consists of two silicon mirrors with a width of $\mathrm{\lambda_{Si}/4}$ that are separated by a distance of $\mathrm{\lambda_{SiO_2}/4}$, where $\mathrm{\lambda_{Si}}$ and $\mathrm{\lambda_{SiO_2}}$ are the wavelengths in silicon and silicon-dioxide, respectively. The schematic of the structure is shown in Fig.~\ref{fig:figure3}a with one new optimization parameter, the reflector distance (D), in addition to the twelve parameters of the fundamental structure shown in Fig.~\ref{fig:figure2}a. Fig.~\ref{fig:figure3}b shows the value of all the thirteen parameters for the newly optimized design. Such a structure demonstrates an increased linear directive gain of 84 directed along $\theta=1^\circ$ and $\varphi=0^\circ$ as seen in Fig.~\ref{fig:figure3}c. This increased gain is attributed to the use of the simple reflector which also becomes evident from the near-field distribution showing an increased upward propagation of the power (see Fig.~\ref{fig:figure3}d). This configuration now exhibits an upward efficiency of 88\%, downward efficiency of 12\%, and a reflection efficiency of 1\% (see Fig.~\ref{fig:figure4}b). Especially, 6.8\% of the optical power is directed into the FOV, which is essentially two times more than that from the fundamental antenna without the reflector. This antenna structure has a footprint of approximately \SI[mode=text]{3.3}{\micro\metre}$\times$\SI[mode=text]{5.2}{\micro\metre}, i.e. similar to the initial structure.
	
	\begin{figure*}[t!]
		\centering
		\includegraphics[width=\textwidth]{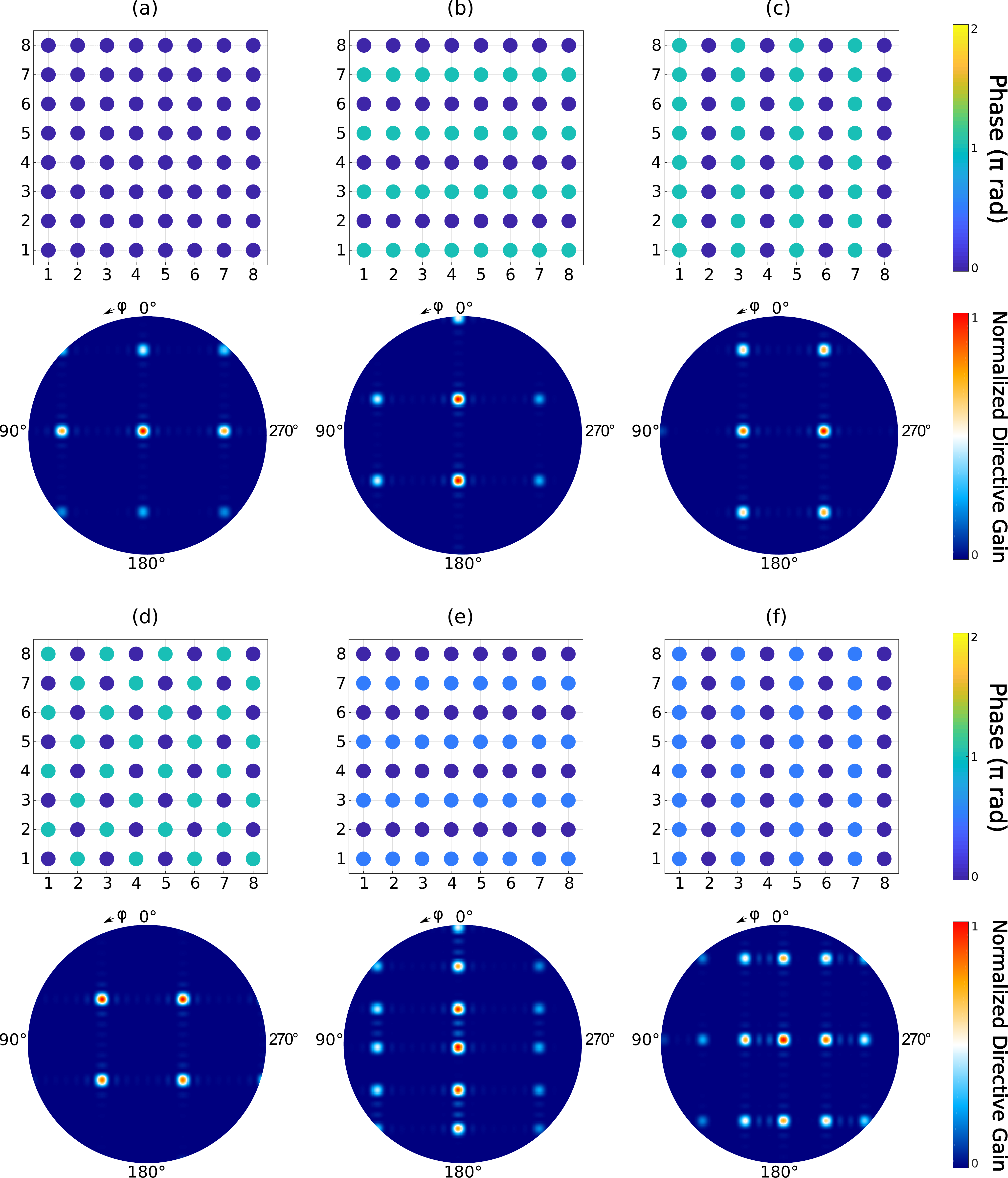}
		\caption{Demonstration of beam steering and beamforming with an $8 \times 8$ array configuration of the antenna with the Bragg reflector. (a) Phase distribution and simulated linear far-field radiation pattern of an uniform phased array. (b-f) Phase distribution and the simulated linear far-field radiation pattern for the main-lobe to be (b) shifted vertically, (c) shifted horizontally, (d) shifted diagonally, (e) split vertically, and (f) split horizontally, respectively.}
		\label{fig:figure6}
	\end{figure*}
	
	To compare the two structures investigated in this work, we plot in Fig.~\ref{fig:figure4}a and b the calculated radiation efficiencies as a function of the wavelength. As it can be seen, both the antennas maintain a good upward efficiency for a broad range of frequencies making them both broadband antennas. Remarkably, the reflector produces not just an increased upward efficiency but also a significantly reduced reflection efficiency of 1\% at the operational wavelength, as depicted by the dashed red curve in Fig.~\ref{fig:figure4}b. Additionally, Fig.~\ref{fig:figure4}c shows the radiation efficiency in the FOV. In comparison to the antenna presented in Ref.\cite{sun2013large1} (grey curve), our optimized antennas without and with the Bragg reflector are able to concentrate around five times and ten times more power in the FOV, respectively. 
	
	Furthermore, the field from a single antenna can be employed for calculation of the electromagnetic far-field radiation of the OPA as 
	\begin{equation}\label{eq1}
		\mathbf{E}_\mathrm{array}(\theta,\varphi)=\mathbf{E}_\mathrm{antenna}(\theta,\varphi) \, \mathrm{AF}(\theta,\varphi),
	\end{equation}
	where $ \mathrm{\mathbf{E}_{antenna}(\theta,\varphi)}$ is the far-field of a single antenna, $\mathrm{AF(\theta,\varphi)}$ is the scalar function representing the array factor, $\mathrm{\mathbf{E}_{array}(\theta,\varphi)}$ is the far-field of the OPA, $\theta$ is the polar, and $\varphi$ is the azimuthal angle of the far-field \cite{balanis2015antenna}. Such arrays are commonly used in beam steering like in radar and LiDAR and conventionally, also the large-scale integration makes it possible to generate complex radiation patterns with high resolution. Our optimized antennas can also be employed in such a 2D-phased array setup for desired pattern generation. The Gerchberg-Saxton algorithm is employed for the pattern synthesis  which utilizes the near-field and far-field intensity distributions to iteratively determine the phase required by each radiating element to generate the desired far-field pattern (in this case, the UPB initials of Paderborn University) \cite{gerchberg1972practical,fienup1978reconstruction}. As the far-field quantity is known, the inverse Fourier transform of this data acts as the initial condition for the algorithm. At any given iteration, the approximate far-field quantity produced by the algorithm is inverse Fourier transformed to provide the magnitude and phase in the near-field. But as the name suggests, the far-field pattern is controlled only by the phase and not the amplitude in a phased array. Therefore, a uniform amplitude distribution of unity is used along with the phases generated by the algorithm for producing the UPB initials. Fig.~\ref{fig:figure5}a and b show the simulated far-field patterns of the $64 \times 64$ array configurations of the optimized antenna with and without the Bragg reflector, respectively. The pink circle highlights the region of the FOV. In agreement with our optimization results it can be seen in Fig.~\ref{fig:figure5}b that the antenna with the reflector generates more power inside the pink circle, manifested in the brighter red of the radiation pattern in comparison to the antenna without the reflector (Fig.~\ref{fig:figure5}a). The original UPB logo used for the pattern generation is shown in the middle and as suggested before, 16 interference orders correspond to 16 repetitions of the logo in each direction. 
	
	Finally, we show the possibility of beam steering and beamforming with our antennas. For this purpose, we considered an $8 \times 8$ array configuration of the antenna with the Bragg reflector. Fig.~\ref{fig:figure6} shows the different phase configurations and the correspondingly generated far-field radiation patterns. In order to have a better visualization of the steering effect, the far-field patterns only show a small region of polar angles, i.e. up to $\theta=10^\circ$. An alternating phase distribution of $0$ and $\pi$ along the rows or columns of the array as shown in Fig.~\ref{fig:figure6}b and c, allows the beam to be steered along the vertical or horizontal direction, respectively. Similarly, using this alternating phase distribution along the rows and columns will shift the position of the beam in both the horizontal and vertical direction, essentially steering the beam along a diagonal as shown in Fig.~\ref{fig:figure6}d. Furthermore, the tightly focused main lobe can also be split vertically or horizontally by using an alternating phase distribution of $0$ and $\pi/2$ along the rows or columns, as shown in Fig.~\ref{fig:figure6}e and f, respectively. 
	
	\section{Conclusion} 
	The dielectric antennas presented in this work were optimized to direct more power into the FOV sized $6.8^\circ \times 6.8^\circ$ and employed in their phased array configuration. Our first optimized antenna demonstrates an upward efficiency of 51\% and 3.2\% of the optical power is directed into the FOV. Our second antenna empowered by the use of a simple, yet efficient Bragg reflector reveals an upward efficiency of approximately 90\% with 6.8\% of the optical power being directed into the FOV, which is ten times more in comparison to the reference antenna \cite{sun2013large1}. This increased efficiency is achieved just with the use of a silicon Bragg reflector. Furthermore, by manipulating the phase distribution of all the antennas comprising the 2D-array, we show the possibility of beam steering and beamforming. We anticipate that these optimized antennas can be easily fabricated and utilized for enhancing the performance of LiDARs and OPAs used in different applications.
    
    \section*{Funding/Acknowledgment}
    The work was funded by the Ministry of Culture and Science of the state of North Rhine-Westphalia via the PhoQC project. Additionally, the authors gratefully acknowledge financial support from the Deutsche Forschungsgemeinschaft (DFG) via TRR142 projects C05 and B06 and the computing time support provided by the Paderborn Center for Parallel Computing (PC$^2$).
    
\bibliography{main}

\begin{thebibliography}{10}

\bibitem{balanis2015antenna}
C.~A. Balanis, {\em Antenna theory: analysis and design}.
\newblock John wiley \& sons, 2015.

\bibitem{ashtiani2019n}
F.~Ashtiani and F.~Aflatouni, ``N $\times$ {N} optical phased array with 2{N}
  phase shifters,'' {\em Optics express}, vol.~27, no.~19, pp.~27183--27190,
  2019.

\bibitem{fenn2000development}
A.~J. Fenn, D.~H. Temme, W.~P. Delaney, and W.~E. Courtney, ``The development
  of phased-array radar technology,'' {\em Lincoln Laboratory Journal},
  vol.~12, no.~2, pp.~321--340, 2000.

\bibitem{sun2013large}
J.~Sun, E.~Timurdogan, A.~Yaacobi, E.~S. Hosseini, and M.~R. Watts,
  ``Large-scale nanophotonic phased array,'' {\em Nature}, vol.~493, no.~7431,
  pp.~195--199, 2013.

\bibitem{benedikovivc2022circular}
D.~Benedikovi{\v{c}}, Q.~Liu, A.~S{\'a}nchez-Postigo, A.~Atieh, T.~Smy,
  P.~Cheben, and W.~N. Ye, ``Circular optical phased array with large steering
  range and high resolution,'' {\em Sensors}, vol.~22, no.~16, p.~6135, 2022.

\bibitem{rabinovich2015free}
W.~S. Rabinovich, P.~G. Goetz, M.~Pruessner, R.~Mahon, M.~S. Ferraro, D.~Park,
  E.~Fleet, and M.~J. DePrenger, ``Free space optical communication link using
  a silicon photonic optical phased array,'' in {\em Free-Space Laser
  Communication and Atmospheric Propagation XXVII}, vol.~9354, p.~93540B,
  International Society for Optics and Photonics, 2015.

\bibitem{neubert1994experimental}
W.~M. Neubert, K.~H. Kudielka, W.~R. Leeb, and A.~L. Scholtz, ``Experimental
  demonstration of an optical phased array antenna for laser space
  communications,'' {\em Applied optics}, vol.~33, no.~18, pp.~3820--3830,
  1994.

\bibitem{blanche2017diffraction}
P.-A. Blanche, L.~LaComb, Y.~Wang, and M.~C. Wu, ``Diffraction-based optical
  switching with mems,'' {\em Applied Sciences}, vol.~7, no.~4, p.~411, 2017.

\bibitem{zhou2015design}
J.~Zhou, J.~Sun, A.~Yaacobi, C.~V. Poulton, and M.~R. Watts, ``Design of 3d
  hologram emitting optical phased arrays,'' in {\em Integrated Photonics
  Research, Silicon and Nanophotonics}, pp.~IT4A--7, Optical Society of
  America, 2015.

\bibitem{poulton2019long}
C.~V. Poulton, M.~J. Byrd, P.~Russo, E.~Timurdogan, M.~Khandaker, D.~Vermeulen,
  and M.~R. Watts, ``Long-range lidar and free-space data communication with
  high-performance optical phased arrays,'' {\em IEEE Journal of Selected
  Topics in Quantum Electronics}, vol.~25, no.~5, pp.~1--8, 2019.

\bibitem{levinson2011towards}
J.~Levinson, J.~Askeland, J.~Becker, J.~Dolson, D.~Held, S.~Kammel, J.~Z.
  Kolter, D.~Langer, O.~Pink, V.~Pratt, {\em et~al.}, ``Towards fully
  autonomous driving: Systems and algorithms,'' in {\em 2011 IEEE intelligent
  vehicles symposium (IV)}, pp.~163--168, IEEE, 2011.

\bibitem{poulton2017coherent}
C.~V. Poulton, A.~Yaacobi, D.~B. Cole, M.~J. Byrd, M.~Raval, D.~Vermeulen, and
  M.~R. Watts, ``Coherent solid-state lidar with silicon photonic optical
  phased arrays,'' {\em Optics letters}, vol.~42, no.~20, pp.~4091--4094, 2017.

\bibitem{bhargava2019fully}
P.~Bhargava, T.~Kim, C.~V. Poulton, J.~Notaros, A.~Yaacobi, E.~Timurdogan,
  C.~Baiocco, N.~Fahrenkopf, S.~Kruger, T.~Ngai, {\em et~al.}, ``Fully
  integrated coherent lidar in 3d-integrated silicon photonics/65nm cmos,'' in
  {\em 2019 Symposium on VLSI Circuits}, pp.~C262--C263, IEEE, 2019.

\bibitem{chung2017monolithically}
S.~Chung, H.~Abediasl, and H.~Hashemi, ``A monolithically integrated
  large-scale optical phased array in silicon-on-insulator cmos,'' {\em IEEE
  Journal of Solid-State Circuits}, vol.~53, no.~1, pp.~275--296, 2017.

\bibitem{abediasl2015monolithic}
H.~Abediasl and H.~Hashemi, ``Monolithic optical phased-array transceiver in a
  standard soi cmos process,'' {\em Optics express}, vol.~23, no.~5,
  pp.~6509--6519, 2015.

\bibitem{van2010two}
K.~Van~Acoleyen, H.~Rogier, and R.~Baets, ``Two-dimensional optical phased
  array antenna on silicon-on-insulator,'' {\em Optics express}, vol.~18,
  no.~13, pp.~13655--13660, 2010.

\bibitem{van2009off}
K.~Van~Acoleyen, W.~Bogaerts, J.~J{\'a}gersk{\'a}, N.~Le~Thomas, R.~Houdr{\'e},
  and R.~Baets, ``Off-chip beam steering with a one-dimensional optical phased
  array on silicon-on-insulator,'' {\em Optics letters}, vol.~34, no.~9,
  pp.~1477--1479, 2009.

\bibitem{doylend2011two}
J.~K. Doylend, M.~Heck, J.~T. Bovington, J.~D. Peters, L.~Coldren, and
  J.~Bowers, ``Two-dimensional free-space beam steering with an optical phased
  array on silicon-on-insulator,'' {\em Optics express}, vol.~19, no.~22,
  pp.~21595--21604, 2011.

\bibitem{doylend2012hybrid}
J.~Doylend, M.~Heck, J.~Bovington, J.~Peters, M.~Davenport, L.~Coldren, and
  J.~Bowers, ``Hybrid iii/v silicon photonic source with integrated 1d
  free-space beam steering,'' {\em Optics letters}, vol.~37, no.~20,
  pp.~4257--4259, 2012.

\bibitem{fatemi2019nonuniform}
R.~Fatemi, A.~Khachaturian, and A.~Hajimiri, ``A nonuniform sparse 2-d
  large-fov optical phased array with a low-power pwm drive,'' {\em IEEE
  Journal of Solid-State Circuits}, vol.~54, no.~5, pp.~1200--1215, 2019.

\bibitem{fatemi2020breaking}
R.~Fatemi, P.~P. Khial, A.~Khachaturian, and A.~Hajimiri, ``Breaking
  fov-aperture trade-off with multi-mode nano-photonic antennas,'' {\em IEEE
  Journal of Selected Topics in Quantum Electronics}, vol.~27, no.~1,
  pp.~1--14, 2020.

\bibitem{guo2013two}
W.~Guo, P.~R. Binetti, C.~Althouse, M.~L. Ma{\v{s}}anovi{\'c}, H.~P. Ambrosius,
  L.~A. Johansson, and L.~A. Coldren, ``Two-dimensional optical beam steering
  with inp-based photonic integrated circuits,'' {\em IEEE Journal of Selected
  Topics in Quantum Electronics}, vol.~19, no.~4, pp.~6100212--6100212, 2013.

\bibitem{sun2021parallel}
C.~Sun, L.~Yang, B.~Li, W.~Shi, H.~Wang, Z.~Chen, X.~Nie, S.~Deng, N.~Ding, and
  A.~Zhang, ``Parallel emitted silicon nitride nanophotonic phased arrays for
  two-dimensional beam steering,'' {\em Optics Letters}, vol.~46, no.~22,
  pp.~5699--5702, 2021.

\bibitem{hulme2015fully}
J.~Hulme, J.~Doylend, M.~Heck, J.~Peters, M.~Davenport, J.~Bovington,
  L.~Coldren, and J.~Bowers, ``Fully integrated hybrid silicon two dimensional
  beam scanner,'' {\em Optics express}, vol.~23, no.~5, pp.~5861--5874, 2015.

\bibitem{resler1996high}
D.~Resler, D.~Hobbs, R.~Sharp, L.~Friedman, and T.~Dorschner, ``High-efficiency
  liquid-crystal optical phased-array beam steering,'' {\em Optics letters},
  vol.~21, no.~9, pp.~689--691, 1996.

\bibitem{mcmanamon1996optical}
P.~F. McManamon, T.~A. Dorschner, D.~L. Corkum, L.~J. Friedman, D.~S. Hobbs,
  M.~Holz, S.~Liberman, H.~Q. Nguyen, D.~P. Resler, R.~C. Sharp, {\em et~al.},
  ``Optical phased array technology,'' {\em Proceedings of the IEEE}, vol.~84,
  no.~2, pp.~268--298, 1996.

\bibitem{sun2013large1}
J.~Sun, E.~Timurdogan, A.~Yaacobi, Z.~Su, E.~S. Hosseini, D.~B. Cole, and M.~R.
  Watts, ``Large-scale silicon photonic circuits for optical phased arrays,''
  {\em IEEE journal of selected topics in quantum electronics}, vol.~20, no.~4,
  pp.~264--278, 2013.

\bibitem{farheen2022broadband}
H.~Farheen, L.-Y. Yan, V.~Quiring, C.~Eigner, T.~Zentgraf, S.~Linden,
  J.~F{\"o}rstner, and V.~Myroshnychenko, ``Broadband optical {T}a$_2$o$_5$
  antennas for directional emission of light,'' {\em Optics Express}, vol.~30,
  no.~11, pp.~19288--19299, 2022.

\bibitem{farheen2022optimization}
H.~Farheen, T.~Leuteritz, S.~Linden, V.~Myroshnychenko, and J.~F{\"o}rstner,
  ``Optimization of optical waveguide antennas for directive emission of
  light,'' {\em JOSA B}, vol.~39, no.~1, pp.~83--91, 2022.

\bibitem{leuteritz2021dielectric}
T.~Leuteritz, H.~Farheen, S.~Qiao, F.~Spreyer, C.~Schlickriede, T.~Zentgraf,
  V.~Myroshnychenko, J.~F{\"o}rstner, and S.~Linden, ``Dielectric travelling
  wave antennas for directional light emission,'' {\em Optics Express},
  vol.~29, no.~10, pp.~14694--14704, 2021.

\bibitem{CST}
Dassault Syst{\'e}mes, "CST Studio Suite,"
  \href{https://www.cst.com}{https://www.cst.com}.

\bibitem{taflove2005computational}
A.~Taflove, S.~C. Hagness, and M.~Piket-May, ``Computational electromagnetics:
  the finite-difference time-domain method,'' {\em The Electrical Engineering
  Handbook}, vol.~3, 2005.

\bibitem{palmer2020diffraction}
C.~Palmer, ``Diffraction grating handbook. mks instruments,'' {\em Inc.: New
  York, NY, USA}, 2020.

\bibitem{roelkens2006high}
G.~Roelkens, D.~Van~Thourhout, and R.~Baets, ``High efficiency
  silicon-on-insulator grating coupler based on a poly-silicon overlay,'' {\em
  Optics Express}, vol.~14, no.~24, pp.~11622--11630, 2006.

\bibitem{fan2007high}
M.~Fan, M.~A. Popovi{\'c}, and F.~X. K{\"a}rtner, ``High directivity vertical
  fiber-to-chip coupler with anisotropically radiating grating teeth,'' in {\em
  Conference on Lasers and Electro-Optics}, p.~CTuDD3, Optica Publishing Group,
  2007.

\bibitem{gerchberg1972practical}
R.~W. Gerchberg, ``A practical algorithm for the determination of plane from
  image and diffraction pictures,'' {\em Optik}, vol.~35, no.~2, pp.~237--246,
  1972.

\bibitem{fienup1978reconstruction}
J.~R. Fienup, ``Reconstruction of an object from the modulus of its fourier
  transform,'' {\em Optics letters}, vol.~3, no.~1, pp.~27--29, 1978.

\end{thebibliography}
\end{document}